\documentclass[a4paper,12pt]{article} %% passando da 12pt a 14pt i caratteri diventano piu'
\usepackage[cmex10]{amsmath}
\usepackage{latexsym,amssymb}
\usepackage[mathscr]{eucal}
\usepackage{amsxtra,amscd,amsthm,upref,layout,color}
\usepackage{calc}
\usepackage[final]{graphicx}
\def\refr#1{{$^{#1}$}}\def\carat{{Carath\'eodory}}
\def\begeq{\begin{equation}}
\def\endeq{\end{equation}}
\def\begdis{\begin{displaymath}}
\def\enddis{\end{displaymath}}
\def\cQ{{\cal Q}}    
  \def\cS{{\cal S}}\def\cV{{\cal V}} 

\def\cQS1{{\cQ_{\cS_1}}}

\def\ie{{\em i.e.}}

\def\ac{\hfill\break\noindent }

\def\href#1{\relax}
\begin{document}
\title{Generalization  of a theorem of Carath\'eodory}
   %{\shorttitle{The algebraic approach}% for neutron scattering}}
\author{%1
{{S. Ciccariello}}${}^a$ {{and A. Cervellino}}${}^b$\\[6mm]
%%%%%%%%
\begin{minipage}[t]{3mm}
    ${}^a$\ 
\end{minipage}
\begin{minipage}[t]{0.9\textwidth}
  \begin{flushleft}
  \setlength{\baselineskip}{12pt}
  {\slshape{\footnotesize{
  Universit\`{a} di Padova, Dipartimento di Fisica ``G. Galilei'' \&{} Unit\`a INFM
  }}}\\{\slshape{\footnotesize{
  Via Marzolo 8, I-35131 Padova, Italy.
  }}}\\{\slshape{\footnotesize{
  E-mail {\upshape{\texttt{salvino.ciccariello@pd.infn.it}}}; Phone +39~049~8277173
  ;\\ Fax +39~049~8277102
  }}}
  \end{flushleft}
\end{minipage}\\[10mm]
\begin{minipage}[t]{3mm}
    ${}^b$\
\end{minipage}
\begin{minipage}[t]{0.9\textwidth}
        \begin{flushleft}
            \setlength{\baselineskip}{12pt}
            {\slshape{\footnotesize{
            Laboratory for Neutron Scattering, PSI Villigen and ETH Zurich
            }}}\\{\slshape{\footnotesize{
            CH-5232 Villigen PSI, Switzerland.
            }}}\\{\slshape{\footnotesize{
            E-mail {\upshape{\texttt{antonio.cervellino@psi.ch}}}; Phone 
            +41~56~3104611;\\ Fax +41~56~3102939
            }}}\\{\slshape{\footnotesize{
            \emph{On leave from:} Istituto di Cristallografia (CNR-IC) 
            }}}\\{\slshape{\footnotesize{
            Via Amendola 122/O, I-70126 Bari, Italy.
            }}}
        \end{flushleft}
        \end{minipage}%\\[10mm]
}%1

%
%\author{
%%\begin{center}
%{{Salvino Ciccariello}}\\[6mm]
%%%%%%%%%
%{\slshape{\footnotesize{
% Dipartimento di Fisica {\em G. Galilei} e Unit\`{a} INFM
%}}}\\
%{\slshape{\footnotesize{
%Via Marzolo 8, I-35131 Padova, Italy.
%}}}\\
%{\slshape{\footnotesize{
%E-mail: {\upshape{\texttt{ciccariello@pd.infn.it}}}; Phone +39~049~8277173
%}}}
%%\end{center}
%}
\date{April 7, 2006}
     % Use \shortauthor to indicate an abbreviated author list for use in
     % running heads (you will need to uncomment it).

\maketitle                        % DO NOT DELETE THIS LINE
%%%! \begin{synopsis}
%%%!\end{synopsis}
\begin{abstract} 
\setlength{\baselineskip}{15truept}
\noindent Carath\'eodory showed that $n$ complex numbers $c_1,\ldots,c_n$ 
can uniquely be written in the form $c_p=\sum_{j=1}^m \rho_j {\epsilon_j}^p$ 
with $p=1,\ldots,n$, where the $\epsilon_j$s are different unimodular complex numbers, 
the $\rho_j$s are strictly positive numbers and integer $m$ never exceeds $n$.
We give the conditions to be obeyed for the former property to hold true if 
the $\rho_j$s are simply required to be real and different from zero. 
It turns 
out that the number of the possible choices of the signs of the $\rho_j$s are 
{at most} equal to the number of the different eigenvalues of the Hermitian  
Toeplitz matrix whose $i,j$-th entry is $c_{j-i}$, where $c_{-p}$ is equal to the 
complex conjugate of $c_{p}$ and 
$c_{0}=0$. 
This generalization is relevant for neutron scattering. Its proof is made 
possible by a lemma - which is an interesting side result - 
that establishes a necessary and sufficient 
condition for the unimodularity of the roots of a polynomial 
based only on the polynomial coefficients. 
%  \\ 
%\data
%\noindent MSC2000: \\
\end{abstract}
\newpage

\setlength{\baselineskip}{25pt}

\section{Introduction}
\carat's theorem\refr{1} states that $n$ complex number $c_1,\ldots,c_n$ 
{ as well as their complex conjugates, respectively denoted by 
$c_{-1},\ldots,c_{-n}$,} can always and uniquely be written as
\begin{equation}\label{1}
c_p=\sum_{j=1}^m \rho_j{\epsilon_j}^p,\quad p=0,\pm 1,\ldots,\pm n,
\end{equation}
with $\rho_j\in\mathbb{R}$, $\rho_j>0$; 
$\epsilon_j\in\mathbb{C}$, $|\epsilon_j|=1$, $\epsilon_j\ne\epsilon_k$ $(j\ne k=1,
\ldots,m)$ { and $c_0$ uniquely determined by the $c_j$s with $j\ne0$}. 
Further,  the $\rho_j$s, the $\epsilon_j$s and $m$ are unique and $m$ obeys to 
$1\le m\le n$. \ac
The practical relevance of this theorem for the inverse scattering problem\refr{2}  
as well as for information theory\refr{3,4} appears evident from the following remark. 
Writing the $\epsilon_j$s as $e^{i2\pi x_j}$ with $0\le x_j<1$, 
the $c_p$'s take the form $\sum_{j=1}^m \rho_j e^{i2\pi x_j p}$ so as 
they can be interpreted as the scattering amplitudes generated by $m$ point 
scatterers (with "charges" $\rho_1,\ldots,\rho_m$ respectively 
located at $x_1,\ldots,x_m$) and relevant to the "scattering vector" 
values $p=0,\pm 1,\ldots,\pm n$.  
One concludes that the theorem of Carath\'eodory allows us 
to determine the positions and the charges from the knowledge of the scattering 
amplitudes $c_1,\ldots,c_n$. Further, it ensures that the solution of 
this inverse problem is unique. However, in the case of neutron scattering\refr{5}, 
the charges of the scattering centers have no longer the same sign. {Nonetheless, 
it has recently been shown\refr{6} that the so-called algebraic approach for 
solving the structure in the case of X-ray scattering from an ideal crystal\refr{2}  
can also be applied to the case of neutron scattering. This result 
suggests that the aforesaid \carat\ theorem can be generalized so as to avoid 
the requirement that the sign of all the scattering charges be positive. In this 
note we show how this generalization is carried out.} \ac
Before proving this statement, we find it convenient to briefly 
sketch the proof of \carat's theorem reported by Grenander and Szeg\"o\refr{7}. 
The proof is based on an enlargement of the set of the $n$ given complex numbers  
$c_p$ to a set containing $(2n+1)$ complex numbers still denoted as $c_p$ 
with $-n\leq p\leq n$, the $c_p$s with negative index being defined as the complex 
conjugates of the given $c_p$s, \ie\  $c_{-p}=\overline{c_p}$ with $p=1,\ldots,n$. 
(Hereafter an overbar will always denote the complex conjugate). 
The remaining value $c_0$, real by  assumption, is determined as follows. Consider 
the $(n+1)\times(n+1)$ matrix $(C)$ with its $(r,s)$th element defined as 
\begin{equation}\label{2}
C_{r,s}=c_{s-r},\quad  r,s=1,\ldots,(n+1). 
\end{equation}
This matrix is a Hermitian Toeplitz matrix\refr{8} and its diagonal elements are 
equal to $c_0$. This value is chosen in such a way that the matrix $(C)$ 
turns out to be singular (\ie\ $\det(C)=0$) and 
the associated bilinear Hermitian form 
\begin{equation}\label{3}
u^{\dag} (C) u \equiv\sum_{r,s=1}^{n+1}{\overline{u_r}}\,C_{r,s}\, u_s
\end{equation}
[where $u$ is an $(n+1)$ dimensional complex vector] non-negative 
(or semi-positive) definite. 
To show that 
$c_0$ can uniquely be determined, one proceeds as follows\refr{3}. Consider the 
$(n+1)\times(n+1)$ matrix $(\hat {C})$ that has its non-diagonal elements equal to 
the correspondent elements of $(C)$ and the diagonal elements equal to zero, \ie, 
with $r,s=1,\ldots,(n+1)$, 
\begin{equation}\label{4}
{\hat C}_{r,s}=
\begin{cases}
c_{s-r},&\text {if $r\ne s$,}\\
0,      &\text {if $r =s$.}
\end{cases}
\end{equation}
This matrix is Hermitian. Then its eigenvalues ($\chi_j,\ j=1,\ldots n+1$), are 
real and can be labeled so as to have $\chi_{_1}\le\ldots\le\chi_{_{n+1}}$. 
Further, they are such 
that $\sum_{j=1}^{n+1}\chi_{_j} =0$ because the trace of $({\hat C})$ is zero. Hence, 
$\chi_{_1}<0$ and $\chi_{_{n+1}}>0$. One immediately realizes that 
{ matrix $(C)$ is obtained by setting} $c_0=-\chi_{_1}>0$ 
so that $(C)=({\hat C})-\chi_{_1}(I)$, $(I)$ being the unit matrix. 
In fact, the matrix $\bigl({\hat C}-\chi_{_1}I\bigr)$ 
is a Toeplitz Hermitian matrix with its diagonal elements equal to 
$(-\chi_{_1})$. The secular equation of this matrix is 
$$\det\bigl((C)-z(I)\bigr)=\det\bigl((\hat C)-\chi_{_1}(I)-z(I)\bigr)
=\det\bigl((\hat C)-(\chi_{_1}+z)(I)\bigr)=0.$$ 
This equation is the same equation that determines the eigenvalues of $({\hat C})$ 
if, instead of variable $z$, we use the shifted variable $z+\chi_{_1}$. Thus, 
the eigenvalues of $(C)$ are: $0=(\chi_{_1}-\chi_{_1})\le (\chi_{_2}-
\chi_{_1})\le\ldots\le
(\chi_{_{n+1}}-\chi_{_1})$ and the matrix $(C)$ is semi-definite positive. Let 
$\mu_{_1}(\ge 1)$ denote the multiplicity of the eigenvalue $\chi_{_1}$ of $({\hat C})$, 
then the rank of $(C)$ is $(n+1-\mu_1)$ and the $m$ present in 
Eq.~(\ref{1}) has the same value, \ie\ $m=(n+1-\mu_{_1})$. Exploiting the 
non-negative definiteness of $(C)$, 
Grenander and Szeg\"o showed that: I) the $\epsilon_j$s are distinct and are the 
roots of the following polynomial equation, {referred to as} {\em resolvent 
equation} {in the following}, 
\begin{equation}\label{5}
P_m(z)={\mathrm D_m}^{-1}\det
\begin{pmatrix}
c_{_0}  &c_{_1}    &\cdots   &c_{_{m-1}}  &c_{_m}\\
c_{_{-1}}   &c_{_0}    &\cdots   &c_{_{m-2}}  &c_{_{m-1}}\\
\hdotsfor{5}\\
c_{_{-m+1}}   &c_{_{-m+2}}    &\cdots   &c_{_{0}}  &c_{_{1}}\\
1             &z              &\cdots   &z^{m-1}   &z^m
\end{pmatrix}=0,
\end{equation}
where ${\mathrm D_m}>0$ is the determinant of the left principal minor 
contained in the first $m$ rows; II) the $\rho_j$s are strictly positive 
and given by
\begin{equation}\label{6}
\rho_j=\frac{1}{{P_m}'(\epsilon_j)}\sum_{p=0}^{m-1}\beta_{j,p}c_p,
\end{equation}
where the prime denotes the derivative and the $\beta_{j,p}$s are the 
coefficients of the following polynomial 
\begin{equation}\label{7}
P_m(z)/(z-\epsilon_j)\equiv \sum_{p=0}^{m-1}\beta_{j,p}z^p;
\end{equation}
and III) that Eq.~(\ref{1}) holds true with $p=-n,(-n+1),\ldots,n$. \ac 
{This theorem, {\em via} (\ref{1}), implies that any non-negative definite 
Hermitian Toeplitz matrix $(C)$, defined by (\ref{2}), uniquely factorizes 
as $(\cV)(Q) (\cV)^{\dag}$ where $(\cV)$ is an $(n+1)\times m$ 
Vandermonde matrix with $\cV_{i,j}\equiv\epsilon_i^{j-1}$, $(\cV)^{\dag}$ its Hermitian 
conjugate  and $(Q)$ an $m\times m$ positive-definite diagonal matrix with 
$Q_{i,j}\equiv\rho_i\delta_{i,j}$. A more general factorization 
was obtained by Ellis and Lay\refr{9}. In their theorem 3.4, they
showed that a Toeplitz matrix of order $(n+1)$ and rank $m\le n$ 
can be factorized as $(\cV)(\Delta) (\cV)^{\dag}$ where $(\cV)$ is a generalized 
confluent Vandermonde matrix generated by the distinct roots of the resolvent equation and 
$(\Delta)$ is a block-diagonal matrix with reversed upper triangular blocks. In proving 
this generalized factorization, the Toeplitz matrix was neither required 
to be Hermitian nor positive/negative semidefinite and this last property - as it 
will appear clear in \S\ 2 - amounts to relax the condition that 
the $\rho_j$s present in (\ref{1}) be strictly positive. However, in proving 
this theorem, it was explicitly assumed that all the roots of the resolvent 
equation were unimodular without stating the conditions that have to be obeyed 
by the Toeplitz matrix elements for the unimodularity property to occur. On the one 
hand, this point makes the generalization not fully proven. On the other hand, the 
physical motivation that underlies our attempt of generalizing \carat's theorem requires 
that Toeplitz matrices are Hermitian as well as a strictly diagonal structure of 
$(\Delta)$.\ac 
The generalization of \carat's theorem presented below meets the last 
two requirements. In particular, in appendix A we report the lemma that 
specifies the 
necessary and sufficient conditions that must be obeyed by the coefficients of 
a polynomial equation for all its roots to lie on the unit circle. We also 
stress that all the so far reported\refr{10,11} theorems, ensuring the unimodularity 
of the roots, are not based on the only knowledge of the polynomial coefficients, 
as our lemma does.} Finally,  appendices B, C and D 
report some properties of the coefficients of the resolvent equation, of Hermitian 
Toeplitz matrices and some numerical illustrations, respectively.
\section{Generalization of \carat's theorem}
The aforesaid choice of $c_0$ is the only one that yields a 
non negative definite matrix $(C)$. It is evident that if we choose 
$c_0=-\chi_{_{n+1}}$ the resulting $(C)$ is a non-positive definite 
matrix and $-(C)$ is a semi-positive definite one. We can apply to 
the latter the same analysis made by Grenander and Szeg\"o with the 
conclusion that 
$$-c_p=\sum_{j=1}^m {\rho'}_j {{\epsilon'}_j}^p,\quad p=0,\pm 1,\ldots,\pm n$$ 
where $m=(n+1-\mu_{_{n+1}})$, $\mu_{_{n+1}}$ denoting the multiplicity 
of $\chi_{_{n+1}}$, the ${\rho'}_j$s are positive and the ${\epsilon'}_j$s 
are distinct unimodular numbers.\ac 
Of course this generalization is quite trivial. The remaining choices are 
however more interesting. In fact, if we choose $c_0=-\chi_j$ with 
$\chi_j\ne \chi_{_1}$ and $\chi_j\ne\chi_{_{n+1}}$, the resulting matrix 
$(C)$ will be indefinite, since some of its eigenvalues are negative, others 
positive and one, at least, equal to zero. The question is now the 
following: does a relation similar to (\ref{1}) hold also true in this case?  
We shall show that the answer is affirmative {iff: i) $m$, the rank 
of $(C)$ smaller than the order $(n+1)$ of $(C)$, is equal to the } {\em 
principal rank} {of $(C)$ and ii)  the conditions reported in the lemma 
discussed in appendix A are fulfilled}. In i) {principal rank means  
the highest of the rank values of all the} {\em strictly principal} {left 
minors of $(C)$ where, by definition, an $(m\times m)$ strictly principal (left) 
minor of $(C)$ is a principal minor formed by $m$} {\em subsequent} {rows 
and columns with indices} $p+1,\ldots,p+m$ and $0\le p\le (n+1-m)$. \ac
Let ${\hat \chi}_{_1}<\ldots<{\hat \chi}_{_{\nu}}$ denote the \emph{distinct} 
eigenvalues of $({\hat C})$ (the matrix with $c_0=0$), $\mu_{_1},\dots,
\mu_{_{\nu}}$ their respective multiplicities (independent from the choice 
of $c_0$) and $n_{\hat \chi}$ and $p_{\hat \chi}$ the number of the 
${\hat \chi}_{l}$s that respectively are negative and positive,  
so that either $\nu=(n_{\hat \chi}+p_{\hat \chi}+1)$ or $\nu=(n_{\hat \chi}+
p_{\hat \chi})$ depending on whether one of the ${\hat\chi}_{l}$s is equal 
to zero or not. If we set 
\begin{equation}\label{8}
c_{l,0}=-{\hat\chi}_{l},\quad N_{l,-}\equiv\sum_{i=1}^{l-1}
\mu_{i}\quad 
{\rm and}\quad N_{l,+}\equiv\sum_{i=l+1}^{\nu}\mu_{i},
\end{equation}
the resulting matrix $(C_{l})$ will have $N_{l,-}$ negative 
eigenvalues, $N_{l,+}$  positive eigenvalues and $\mu_{l}$ 
eigenvalues equal to zero so that its rank is $m_{l}=(n+1-\mu_{l})$. 
We shall prove that the following relations 
\begin{eqnarray}
c_p&=&\sum_{j=1}^{m_{l}} \rho_{l,j} {\epsilon_{l,j}}^p,
\quad p=\pm 1,\ldots,\pm n,\label{9}\\
c_{l,0}&=&\sum_{j=1}^{m_{l}} \rho_{l,j},\quad p=0\nonumber
\end{eqnarray} 
hold true { if $m_{l}$ is equal to the principal rank of $(C_{l})$} 
and the conditions stated in the lemma are obeyed. 
In (\ref{9}) the value of $m_{l}$ has been just defined, the 
$\epsilon_{l,j}$s are the distinct unimodular roots of a polynomial 
equation of degree $m_{l}$ uniquely determined by the $c_p$s and 
$c_{l,0}$ and, finally, the number of the positive and negative 
$\rho_{l,j}$s respectively is $N_{l,+}$ and $N_{l,-}$. \ac
{ It is remarked that we can construct $\nu$ different matrices 
$(C_{l})$ from a given $({\hat C})$ by setting $c_{l,0}=-
{\hat\chi}_{l},\ l=1,\ldots,\nu$. {Among these matrices only 
those which obey i) and ii)} allow us to write $c_{l,0}$ 
and the $c_p$s, with $p=\pm1,\ldots,\pm n$, in the form (\ref{9}). 
Hence, if the number of these matrices is denoted by $\mu$, one 
concludes that we have  $\mu$ different ways for writing the $n$ 
complex numbers $c_{_1},\ldots,c_{_n}$ in the form (\ref{9}). 
In general $\mu$ is such that $2\le \mu\le\nu$.}  Clearly this 
result generalizes \carat's theorem and we pass now to prove it. \ac 
The proof must be achieved by a procedure different from that followed 
by Grenander and Szeg\"o because matrices $(C_{l})$ are no longer 
non-negative definite. Since our attention will focus on a particular 
$(C_{l})$, for notational simplicity we shall omit index $l$  
in the following considerations. Assume that Eqs~(\ref{9}) are fulfilled 
so that
\begin{equation}\label{10}
C_{r,s}=c_{s-r}=\sum_{j=1}^m\rho_j\,\epsilon_j^{s-r},\quad r,s=1,\ldots,(n+1),
\end{equation}
with
\begin{equation}\label{11}
\rho_j\in \mathbb{R},\quad \rho_j\ne 0,\quad \epsilon_j\in\mathbb{C},
\quad|\epsilon_j|=1, 
\quad \epsilon_j\ne\epsilon_i\ \ {\rm if}\ \ j\ne i,\quad j,i=1,\ldots,m.
\end{equation}
Then $C_{r,s}=\overline{C_{s,r}}$ and the Hermiticity of $(C)$ is ensured. 
Introduce now two further matrices: $(V)$ and $(Q)$ such that 
\begin{eqnarray}
V_{j,r}&\equiv& \epsilon_j^{r-1},\quad j=1,\ldots,m,\quad r=1,\ldots,(n+1),
\label{12}\\
Q_{j,i}&\equiv&\rho_j\delta_{j,i},\quad j,i=1,\ldots,m.\label{13}
\end{eqnarray}
The first is an $m\times(n+1)$ rectangular Vandermonde matrix and the second 
a diagonal $m\times m$ matrix. As the $\epsilon_j$s are distinct, the rows of 
$(V)$ are linearly independent. From the unimodularity of the $\epsilon_j$s 
and (\ref{10}) immediately follows that 
\begin{equation}\label{14}
\sum_{i,j=1}^m {V^{\dag}}_{r,i}Q_{i,j}V_{j,s}=
\sum_{i,j=1}^m {\overline{V}_{i,r}}\rho_i\delta_{i,j}V_{j,s}=
%\sum_{i}^m {\overline {V}_{i,r}}\rho_i V_{i,s}=
\sum_{i=1}^m \rho_i \epsilon_i^{s-r}=C_{r,s}
\end{equation}
so that one has 
\begin{equation}\label{15}
(C)=(V)^{\dag}(Q)(V).
\end{equation}
Consider now the $p\times p$ minor of $(C)$ formed by the rows 
$r_1<\cdots<r_{p}$ and the columns $s_1<\cdots<s_{p}$. By Bezout's 
theorem\refr{12,13} one finds that 
\begin{equation}\label{16}
{\det\bigl(C
\overset{\overset{s_1,\cdots,s_p}{}}{\underset{r_1,\cdots,r_p}{}}
\bigr)} = 
\sum_{1\le i_1<i_2<\cdots<i_p\le m}\bigl[\prod_{j=1}^p\rho_{i_{j}} \bigr]
{\det\bigl(V
\overset{\overset{s_1,\cdots,s_p}{}}{\underset{i_1,\cdots,i_p}{}}
\bigr)
\det\bigl({\overline V}
\overset{\overset{r_1,\cdots,r_p}{}}{\underset{i_1,\cdots,i_p}{}}
\bigr) }%\quad\quad\quad\quad(16)
\end{equation} 
where, {adopting Gantmacher's notation}\refr{12}, the lower and upper 
indices inside each determinant symbol denote the rows and the columns 
of the considered minors of $(C)$, $(V)$ and $({\overline V})$. This 
expression makes it clear that the determinant of any minor of $(C)$ 
of order $p>m$ is equal to zero because the order of $(Q)$ is $m$\refr{14}. 
For this reason the rank of $(C)$ cannot exceed $m$. At the same 
time, if $p=m$, any  $(m\times m)$ { strictly} principal minor 
${\det\bigl(C
\overset{\overset{q+1,\cdots,q+m}{}}{\underset{q+1,\cdots,q+m}{}}
\bigr)}$ (with $0\le q\le n+1-p$) of $(C)$ will have determinant equal to 
\begin{equation}\label{17}
\bigl[\prod_{j=1}^m \rho_j\bigr]\, \Bigl|\prod_{1\le i\le j\le m}
(\epsilon_j-\epsilon_i)\Bigr|^2.
\end{equation}
{In fact, if $p=m$, the sum present in (\ref{16}) involves a single term 
and, due to (\ref{12}), in $\det\bigl(V\overset{\overset{q+1,\cdots,q+m}{}}
{\underset{1,\cdots,m}{}}\bigr)$ we can factorize $\epsilon_1^{q}$ in 
the first row, $\epsilon_2^{q}$ in the second row and so on. The 
remaining matrix is a Vandermonde matrix so that }
\begin{equation}\nonumber
\det\bigl(V\overset{\overset{q+1,\cdots,q+m}{}}
{\underset{1,\cdots,m}{}}\bigr)=\prod_{l=1}^m {\epsilon_l}^q 
\underset{1\le i\le j\le m}{\prod}(\epsilon_j-\epsilon_i).
\end{equation}
{In the same way one shows that }
\begin{equation}\nonumber
\det\bigl({\bar V}\overset{\overset{q+1,\cdots,q+m}{}}
{\underset{1,\cdots,m}{}}\bigr)=\prod_{l=1}^m {\overline{\epsilon_l}}^q 
\underset{1\le i\le j\le m}{\prod}({\overline{\epsilon_j}}-
{\overline{\epsilon_i}}).
\end{equation}
Finally, the unimodularity of the $\epsilon_j$s yields Eq.~(\ref{17}). 
In this way, it has been shown that, if the elements of $(C)$ have the 
expression reported on the right hand side (rhs) of (\ref{10}) and 
conditions (\ref{11}) are obeyed, matrix $(C)$ has rank $m$ and each of 
its principal minors formed by $m$ subsequent rows is nonsingular. 
{In this way the rank of $(C)$ is equal to its principal rank and the 
necessity of condition i) is now clear\refr{15}}. Since these minors 
are equal, the first $m$ rows (or columns) of $(C)$  are linearly 
independent. One recalls that the polynomial equation having its 
roots equal to the $m$ $\epsilon_j$s, can be written as 
\begin{equation}\label{18}
P_m({z})\equiv\prod_{j=1}^m (z-\epsilon_{j})\equiv
\sum_{l=0}^m a_{l} z^{l}=0
\end{equation}
with
\begin{equation}\label{19}
a_{l}= 
(-)^{m-l}\sum_{1\le i_1<\cdots<i_{{m-l}}\le {m}}
\epsilon_{i_1}\cdots\epsilon_{i_{{m-l}}},\quad l=0,
\ldots,(m-1)
\end{equation}
and $a_m=1.$ 
{ From (\ref{19}) and the unimodularity of the $\epsilon_{l}$s follows that 
$a_0$ is unimodular and, therefore, is different from zero.} 
From the condition $P_m(\epsilon_{i})=0$ follows that 
$\epsilon_{i}^m=-\sum_{l=0}^{m-1}a_{l} \epsilon_{i}^{l}$. 
After multiplying the latter by ${\epsilon_{i}}^{j}$, 
with $j\in {\mathbb{Z}}$, and setting $q=m+j$, one finds that 
\begin{equation}\label{20}
\epsilon_{i}^{q}=-\sum_{l=0}^{m-1}a_{l} \epsilon_{i}^{l+q-m},
\quad q\in {\mathbb{Z}}\quad {\rm and}\quad i=1,\ldots,m.
\end{equation}
The substitution of these relations in (\ref{10}) yields 
\begin{equation}\label{21}
c_{s-r}=-\sum_{i=1}^m\rho_{i}\sum_{l=0}^{m-1}a_{l} 
\epsilon_{i}^{l+s-r-m}=
%-\sum_{l=0}^{m-1}a_{l}\sum_{i=1}^m\rho_{i}
%\epsilon_{i}^{l+s-r-m}=
-\sum_{l=0}^{m-1}a_{l} c_{s-r-(m-l)}.
\end{equation}
Taking $(s-r)=1,2,\ldots,m$, one obtains the following system of linear 
equations
\begin{equation}\label{22}
\begin{array}{l l l l c l c l}
& c_{0}\ \ a_0 & +c_{1}\ a_1 &+ c_{2}\ a_2+&.. &+c_{m-1}\,a_{m-1}& = &-c_{m}\\
& c_{-1}\ a_0& +c_{0}\ a_1&+ c_{1}\ a_2+&.. &+c_{m-2}\,a_{m-1} & = & -c_{m-1} \\
& .       & .     &.    &.  &...        & . & .\\
& c_{-m+1}\, a_0& + c_{-m+2}\,a_1&+c_{-m+3}\, a_2+&.. &+c_{0}\,a_{m-1} & = & -c_{1}.
\end{array}
\end{equation}
This uniquely determines coefficients $a_0,\ldots,a_{m-1}$ of (\ref{18}) 
because the determinant of the coefficients in (\ref{22}) is different from 
zero and equal to ${\mathrm D_m}$, defined below Eq. ({5}). Recalling that 
$a_{m}=1$, resolvent equation (\ref{18}) is fully determined and formally 
coincides with Grenander \& Szeg\"o's Eq.~(\ref{5}).\ac
The solution of (\ref{22}) is 
\begin{align}
a_{l}&=
{(-)^{m-l}\det\bigl(C\overset{\overset{1,\cdots,l,l+2,\cdots,m+1}{}}
{\underset{1,\cdots,m}{}}\bigr)}/{\mathrm D},\ \quad l=0,\ldots,m\label{26}\\
{\mathrm D}&\equiv \det\bigl(C\overset{\overset{1,\cdots,m}{}}
{\underset{1,\cdots,m}{}}\bigr)\ne 0,\label{27}
\end{align}
where {in (\ref{26}) we let $l$ assume the value $m$ since it yields $a_m=1$ 
as reported below (\ref{19}).} 
{Further, setting $l=0$ in Eq.(\ref{26}) and using (\ref{B6}) (with $n=m+1$, $p=1$ 
and $q=2$), one finds 
\begin{equation}\label{27bis}
a_0={(-)^{m}\det\bigl(C\overset{\overset{2,\cdots,m+1}{}}
{\underset{1,\cdots,m}{}}\bigr)}/{\mathrm D}=(-)^m e^{i\theta_{_{1}}}.
\end{equation}
In this way one recovers that $a_0$ is unimodular as already reported below 
(\ref{19}). As discussed in appendix C, coefficients (\ref{26}) are partly 
related by the relations ${\overline {a_l}}=a_{m-l}/a_0$, $l=0,\ldots,m$. These 
imply that 
the roots of (\ref{18}) obey to ${\overline{\epsilon_{i_{j}}}}=1/\epsilon_j$ 
with $j=1,\ldots,m$ and $i_1,\ldots,i_m$ equal to a permutation of 
$\{1,\ldots,m\}$. This result ensures the unimodularity only for those 
roots with ${i_{j}}=j$. The unimodularity of all the roots is ensured 
by the further conditions stated in the lemma of appendix A. Thus,  
the roots of (\ref{18}) are 
unimodular if the new $(m+1)\times (m+1)$ Toeplitz matrix $(\cS)$, defined 
as 
\begin{align}
\cS_{r,s}&\equiv \sigma_{s-r},\quad r,s=1,\ldots,m+1,\label{27ter}\\
\sigma_p&\equiv\sum_{j=1}^m{\epsilon_j}^p, p=0,\pm1,\pm2,\ldots,\pm m\nonumber,
\end{align} 
is nonnegative definite and has rank $m$. 
Quantities $\sigma_p$ are easily and uniquely determined by Eq.s (\ref{A3}) 
and (\ref{A4}) in terms of the $a_p$s given by (\ref{26}). 
Thus, after checking that the determinants of the left principal minors,   
contained in the first $p$ rows of $(\cS)$, are strictly positive for $p=1,\ldots,m$  
and equal to zero for $p=m+1$, the unimodularity of all the roots of (\ref{18}) 
is ensured. After solving the resolvent equation, the $\rho_j$s {can be determined 
by Eq.s (\ref{6}) and (\ref{7}) since these also 
apply in the case of non-definite $(C)$.} Alternatively, the $\rho_j$s 
can be determined solving the system of $m$ linear equations 
\begin{equation}\label{29}
\sum_{j=1}^{m} {\epsilon_{j}}^p \rho_{j}=c_p,\ \quad p=0,\ldots,(m-1) 
\end{equation}
that follow from (\ref{9}). [For notational simplicity, we still omit index 
$l$ present in the definition of $c_0$]. These equations can also be written as 
\begin{equation}\label{30}
\sum_{j=1}^{m} V^T_{p+1,j} \rho_{j}=c_p,\ \quad p=0,\ldots,(m-1) 
\end{equation}
where $(V^T)$ is the transpose of the $m\times m$ upper left principal minor
of matrix $(V)$ defined by (\ref{12}). The formal solution of (\ref{30}) is 
\begin{equation}\label{31}
\rho_{j}=\sum_{p=0}^{m-1} {(V^T)^{-1}}_{j, p+1} c_p,\quad j=1,\ldots,m,
\end{equation}
since $(V^T)$ is certainly non singular.\ac
Finally it must be proven that the numbers of the $\rho_j$s that turn out to be 
positive or negative are respectively equal to $N_{+}$ and $N_-$ (again we omit 
index $l$). To this aim consider the Hermitian bilinear form 
\begin{equation}\label{32}
{\mathbb C}_2[u]\equiv\sum_{r,s=1}^{n+1} {\overline {u_r}}C_{r,s}u_s,\ 
\quad u_s\in{\mathbb C}.
\end{equation}
By Eq~(\ref{15}) this is immediately expressed in terms of the diagonal form 
\begin{equation}\label{33}
{\mathbb C}_2[u]= \sum_{p=1}^m {\overline {v_p[u]}}\rho_p v_p[u]
\end{equation}
where 
\begin{equation}\label{34}
v_p[u]\equiv \sum_{s=1}^{n+1} V_{p,s}u_s,\quad\ p=1,\cdots,m.
\end{equation}
At the same time, since $(C)$ is Hermitian, it can  be diagonalized by a unitary 
transformation $(U)$ and written as 
\begin{equation}\label{35}
(C)=(U)^{\dag}(\chi)(U)
\end{equation} 
where
\begin{equation}\label{36}
(\chi)_{r,s}=(\chi_r-{\hat{\chi}}_{l})\delta_{r,s},\ \quad r,s=1,\ldots,(n+1)
\end{equation}
with the $\chi_r$s equal to the eigenvalues of $({\hat C})$. As discussed at the 
beginning of this section, $\mu_{l}$ of the $(\chi_r-{\hat{\chi}}_{l})$s are equal to 
zero, $N_{l,+}$ are positive and $N_{l,-}$ negative. {Therefore we can compact 
$(U^{\dag})$ by eliminating $\mu_{l}$ columns whose index correspond to 
the rows of $(\chi)$ containing the zero eigenvalues and, subsequently, $(\chi)$ 
eliminating the rows and the columns  containing the zero eigenvalues. 
Hereinafter $(U)$ shall be an $m\times(n+1)$ rectangular matrix with orthonormal rows 
and $(\chi)$ an $m\times m$ diagonal and nonsingular matrix. We set
\begin{equation}\label{38}
w_p\equiv \sum_{s=1}^{n+1}U_{p,s}u_s,\ \quad p=1,\ldots,m
\end{equation}
and consider the $w_p$ as the arbitrary independent variables. 
Using Eq.~(\ref{15}) we can write
\[(U)^{\dag}(\chi)(U)=(C)=(V)^{\dag}(Q)(V).\]
The row-spaces of $(U)$ and $(V)$ necessarily coincide with the $(U)$ 
and $(V)$ right image spaces that in turn coincide with the eigenspace 
of $(C)$ associated to the eigenvalue zero. 
There exists then a non-singular $m\times m$ matrix $(R)$ such that
\[(V)=(R)(U).\]
Now, for any complex $m$-tuple 
$\pmb{w}=(w_{1},\ldots,w_{m})$, we have 
$\pmb{w}^{\dag}(\chi)\pmb{w}=\pmb{w}^{\dag}(R)^{\dag}(Q)(R)\pmb{w}$ 
which leads to $(\chi)=(R)^{\dag}(Q)(R)$. 
Thus $(\chi)$ and $(Q)$ are related by a congruence and Sylvester's 
inertia law\refr{17} applies, and the number of positive (negative) 
$\rho_p$s coincides with the number of positive (negative) 
$(\chi_p-{\hat \chi}_{l})$s. In this way the generalization of the 
Carath\'eodory theorem is complete.}
\section{Conclusion}
Summarizing, given $n$ complex numbers $c_1,\ldots,c_n$ one considers the
Hermitian Toeplitz matrix $({\hat C})$ defined by (\ref{4}). One evaluates its 
distinct eigenvalues, denoted by ${\hat \chi}_1<\ldots<{\hat \chi}_{\nu}$ with 
mutiplicities $\mu_1,\ldots,\mu_{\nu}$. 
Setting $(C_l)\equiv({\hat C})-{\hat {\chi}}_l(I)$ 
with $l=1,\ldots,\nu$, the resulting matrices with $l\ne 1$ and $l\ne\nu$
are indefinite. For each of these $l$ values, the complex numbers
$-{\hat \chi}_l,c_1,\ldots,c_n$ also can uniquely be written in the form (\ref{1})
with $m_l=(-\mu_l+\sum_{q=1}^{\nu}\mu_q)$ iff i)  the rank and the principal
rank of $(C_l)$ are equal [its value turns out to be equal to $m_l$]  and ii)
the $(m_l+1)\times(m_l+1)$ matrix $(\cS_l)$ [defined by
(\ref{27ter}), (\ref{A2}), (\ref{A3}) and (\ref{26})] is 
non-negative definite and has rank $m_l$. {In proving these results 
it is essential to know the conditions that must be obeyed by the coefficients 
of a polynomial equations for all its distinct roots to lie on the unit circle. The 
answer to this problem is given by the lemma reported in appendix A.}\ac
As last remark we observe that, in theorems 2.2 and 3.4 of Ellis and Lay\refr{9}, the 
assumption that the resolvent has unimodular roots can be removed  by the aforesaid lemma. 
It can be substituted with the constructive requirements that: a) the coefficients 
of the resolvent equation obey Eq.~(\ref{A1bis}) if the given Toeplitz matrix $({\cal T})$ 
is not Hermitian (oppositely, the condition is already fulfilled), b) if the discriminant of 
the resolvent equation is equal to zero, one algebraically eliminates\refr{18,19} all the 
multiple roots from the resolvent obtaining the lowest degree resolvent equation [\ie\ 
the equation with roots equal to all the distinct roots of the outset resolvent], 
c) from the coefficients of the (new) resolvent equation one constructs matrix $(\cS)$ 
defined by (\ref{27ter}) and one checks its positive definiteness. In the only affirmative 
case the Ellis-Lay generalized factorization of $({\cal T})$ is possible. This reduces to 
\carat's generalized one if $({\cal T})$ is Hermitian and the outset resolvent has no 
multiple roots [\ie\ step b) is not required].

\vfill\eject
\appendix
\section{Unimodular roots' conditions}
{We shall now prove a lemma that states the necessary and sufficient condition for 
all the zeros of a polynomial equation with complex coefficients lie on the unit 
circle. The so far known theorems that ensure such property leans upon the existence of other 
polynomial with unimodular roots\refr{10,11}, while the following lemma only involves the 
coefficients of the given polynomial.}\ac
In appendix  B it is shown that, if the coefficients $a_{l}$ of 
the $N$th degree polynomial equation 
\begin{equation}\label{A1}
P_N(z)\equiv\underset{1\le j\le N}\prod(z-\epsilon_j)=\sum_{l=0}^Na_{l}z^{l}=0,
\end{equation}
obey the following conditions 
\begin{equation}\label{A1bis}
a_N\equiv 1,\quad|a_0|=1\quad{\mathrm and}\quad {\overline{a_m}}=a_{N-m}/a_0
\ {\mathrm {for}}\ \ 
m=0,\ldots,N,
\end{equation}
the roots of the equation are such that 
${\overline {\epsilon_{j}}}=1/{{\epsilon_{i_j}}}$ for $j=1,\ldots,m$ and 
$i_1,\ldots,i_m$ equal to a permutation of $\{1,\ldots,m\}$.  
The unimodularity of all the roots being not assured by this property, 
it is natural to ask: which are the further conditions to be obeyed by 
the $a_l$s for all the roots to lie on the unit circle? \ac
The answer is given in the lemma reported later and based on \carat's theorem.\ac
{We first observe that it is not restrictive to assume -as we do below - that 
the roots of Eq. (\ref{A1}) are distinct because possible multiple roots can 
algebraically be eliminated\refr{18,19}.}  
Consider the following symmetric functions of the roots of (\ref{A1}) 
\begin{equation} \label{A2}
\sigma_p \equiv \sum_{j=1}^N {\epsilon_j}^p\quad p=0,\pm 1,\pm 2,\ldots.
\end{equation}
They exist for negative $p$ integers because $a_0\ne 0$. 
For non-negative $p$s  the $\sigma_p$s are uniquely determined 
from the coefficients of (\ref{A1}) by the 
following relations (see, {\em e.g.}, Ref. [18], Chap. XIII)
\begin{equation}\label{A3}
\begin{array}{cccccc}
 &  &  & &    \hfill \sigma_0&=\quad N\\
& & & &       \hfill a_N \sigma_1&= -a_{N-1}\\
 & & &a_N \sigma_2  &+a_{N-1}\sigma_1&= -2 a_{N-2}\\
 & &a_N\sigma_3&+a_{N-1}\sigma_2 &+a_{N-2}\sigma_1&= -3 a_{N-3}\\
\hdotsfor{5}&=\quad ..\\
a_N\sigma_{N-1}+&\ldots &+a_4 \sigma_3&+a_3\sigma_2&+a_2\sigma_1&= -(N-1)a_1\\
a_N\sigma_{p+N}+&\ldots &+a_2\sigma_{p+2}&+a_1 \sigma_{p+1}&+a_0 \sigma_{p}&= 0,
\quad p=0,1,\ldots
\end{array}
\end{equation}
{Owing to the condition ${\overline {\epsilon_{j}}}=1/{{\epsilon_{i_j}}}$, from 
(\ref{A2}) follows that }
\begin{equation}\label{A4}
\sigma_{-p} ={\overline \sigma_{p}}\quad p=1,2,\ldots,
\end{equation}
and the last of relations (\ref{A3}) holds also true for 
negative $p$ integers. In fact, the complex conjugate of this relation by 
(\ref{A1bis}) becomes 
\begin{align}
a_0({\overline{a_N}}\ {\overline{\sigma_{p+N}}}&+\ldots 
+{\overline{a_2}}\ {\overline{\sigma_{p+2}}}+
{\overline{a_1}}\ {\overline{ \sigma_{p+1}}}+
{\overline{a_0}}\ {\overline{ \sigma_{p}}})= \nonumber\\
a_0\sigma_{-p-N} &+\ldots 
+a_{N-2}\sigma_{-p-2} +
{a_{N-1}}\sigma_{-p-1}+a_N\sigma_{-p}=0, \label{A5bis}
\end{align}
and the statement is proven. The previous considerations show 
that all the $\sigma_p$s are known in terms of $a_0,\ldots,a_N$. \ac
Introduce now the $(N+1)\times(N+1)$ Hermitian Toeplitz matrix $(\cS)$ 
having its $(i,j)$th element defined as 
\begin{equation}\label{A6}
\cS_{i,j}\equiv\sigma_{j-i},\quad i,j=1,\ldots,N+1.
\end{equation}
Assume first that the $\epsilon_j$s are unimodular (and distinct) and, similarly 
to (\ref{12}), introduce a further $N\times(N+1)$ matrix $(\cV)$ with 
$\cV_{r,s}\equiv{\epsilon_r}^{s-1}$. The assumed properties of the 
$\epsilon_j$s ensure that $(\cS)=(\cV^{\dag})(\cV)$, that $\det(\cS)\ne 0$ 
and that the rank of $(\cV)$ is $N$. The three properties in turn imply that 
$(\cS)$ is a non negative definite matrix of rank $N$. Then, from \carat's 
theorem follows that the $\sigma_p$s can uniquely be written as
\begin{equation}\label{A6bis}
\sigma_p=\sum_{j=1}^N\tau_j{\omega_j}^p,\ \ p=0,\pm1,\ldots,\pm N
\end{equation}
with the $\omega_j$s unimodular, distinct and roots of the resolvent 
equation generated by matrix $(\cS)$, \ie
\begin{equation}\label{A6ter}
Q_N(z)={\mathrm {\Delta}_N}^{-1}\det
\begin{pmatrix}
\sigma_{_0}  &\sigma_{_1}    &\cdots   &\sigma_{_{N-1}}  &\sigma_{_N}\\
\sigma_{_{-1}}   &\sigma_{_0}    &\cdots   &\sigma_{_{N-2}}  &\sigma_{_{N-1}}\\
\hdotsfor{5}\\
1             &z              &\cdots   &z^{N-1}   &z^N
\end{pmatrix}=0
\end{equation}
(here $\Delta_N$ denotes the determinant of the $N\times N$ upper left principal 
minor of $(\cS)$). The comparison of (\ref{A6bis}) with (\ref{A2}) and the 
uniqueness of the \carat\ decomposition imply that 
$\tau_1=\ldots=\tau_N=1$ and $\{\omega_1,\ldots,\omega_n\}=
\{\epsilon_1,\ldots,\epsilon_N\}$. From the last 
follows that $Q_N(z)=P_N(z)$. \ac 
At this point we can state the lemma:\ac 
{\em  the roots of an $N$ degree polynomial equation $P_N(z)=0$ are unimodular and 
distinct iff its coefficients $a_l$, besides obeying conditions (\ref{A1bis}), are 
such that matrix $(\cS)$, defined by (\ref{A6}), is non-negative definite 
and has rank N.}\ac
\# The necessity of the lemma has already been proven. To prove its sufficiency 
one has to show that the properties that the rank of $(\cS)$ is $N$ 
and that $(\cS)$ is non-negative definite ensure that the roots of $P_N(z)=0$ 
are distinct and unimodular, respectively. 
%$Q_N(z)=P_N(z)$. 
In fact, the first property implies that $\Delta_N\ne 0$. 
From definition (\ref{A2}) and property (\ref{A4}) follows  that 
\begin{align}
\Delta_{N}&={\mathrm \det
\begin{pmatrix}
N  &\sum_{j=1}^N\epsilon_{j}&\sum_{j=1}^N\epsilon_{j}^2&\cdots&
\sum_{j=1}^N\epsilon_{j}^{N-1}\\
\sigma_{_{-1}}   &\sigma_0&\sigma_1&\cdots   &\sigma_{_{N-2}}  \\
\hdotsfor{5}\\
\sigma_{_{-N+1}}   &\sigma_{_{-N+2}}&\sigma_{_{-N+3}}&\cdots   &\sigma_{_{0}} 
\end{pmatrix}}\label{A20}\\
\quad&=\sum_{j=1}^N {\mathrm \det}
\begin{pmatrix}
1  &\epsilon_{j}&{\epsilon_{j}}^2&\cdots   &{\epsilon_{j}}^{N-1}\\
\sigma_{{-1}}   &\sigma_0&\sigma_1&\cdots   &\sigma_{_{N-2}}  \\
\hdotsfor{5}\\
\sigma_{_{-N+1}}&\sigma_{_{-N+2}}&\sigma_{_{-N+3}}&\cdots   &\sigma_{_{0}} 
\end{pmatrix}\nonumber\\  
\quad&=\sum_{1\le j_1,\ldots,j_N\le N}{\mathrm \det}
\begin{pmatrix}
1  &\epsilon_{j_{1}}&{\epsilon_{j_{1}}}^2&\cdots   &{\epsilon_{j_{1}}}^{N-1}\\
{\epsilon_{j_{2}}}^{-1}&1&{\epsilon_{j_{2}}}&\cdots   &{\epsilon_{j_{2}}}^{N-2}\\
\hdotsfor{5}\\
{\epsilon_{j_{N}}}^{-N+1}&{\epsilon_{j_{N}}}^{-N+2}&{\epsilon_{j_{N}}}^{-N+3}&\cdots&1
\end{pmatrix}.\nonumber
\end{align}
The last expression can also be written as 
\begin{equation}\label{A21}
\sum_{1\le j_1,\ldots,j_N\le N}\frac{1}{\epsilon_{j_{1}}^0\epsilon_{j_{2}}^1\ldots
{\epsilon_{j_{N}}}^{N-1}}
{\mathrm \det}
\begin{pmatrix}
1  &\epsilon_{j_{1}}&\epsilon_{j_{1}}^2&\cdots   &{\epsilon_{j_{1}}}^{N-1}\\
1&{\epsilon_{j_{2}}}&\epsilon_{j_{2}}^2&\cdots   &{\epsilon_{j_{2}}}^{N-1}\\
\hdotsfor{5}\\
1&{\epsilon_{j_{N}}}&\epsilon_{j_{N}}^{2}&\cdots&{\epsilon_{j_{N}}}^{N-1}
\end{pmatrix}.
\end{equation}
Within the sum the only terms with $j_1\ne j_2 \ne\ldots\ne j_N$ can 
differ from zero. In other words, the 
possible values of $\{j_1,\ldots,j_N\}$ correspond to the possible permutations of 
$\{1,\ldots,N\}$. The values of the corresponding determinants are 
$(-)^P\prod_{1\le i<j\le N}(\epsilon_j-\epsilon_i)$ where $P$ is the number of 
the transpositions required for passing from $\{j_1,\ldots,j_N\}$ to $\{1,\ldots,N\}$. 
One concludes that 
\begin{equation}\label{A22}
\Delta_N=\prod_{1\le i<j\le N}(\epsilon_j-\epsilon_i)(1/\epsilon_j-1/\epsilon_i).
\end{equation}
Thus, $\Delta_N\ne 0$ ensures that the roots of $P_N(z)=0$ are distinct. 
We show now that the resolvent of $(\cS)$, \ie\ Eq. (\ref{A6ter}), coincides 
with $P_N(z)$. In fact, $Q_N(z)$ can be written as 
$Q_N(z)\equiv \sum_{p=0}^N q_p z^p=0$ with 
\begin{equation} \label{A23}
q_p\equiv \frac{(-1)^{N+p}}{\mathrm {\Delta}_N}{\mathrm \det
\begin{pmatrix}
\sigma_{_{0}}  &\cdots&\sigma_{_{p-1}}&\sigma_{_{p+1}}&\cdots 
&\sigma_{_{N-1}}&\sigma_{_{N}}\\
\sigma_{_{-1}}&\cdots&\sigma_{_{p-2}}&\sigma_{_{p}}&\cdots
&\sigma_{_{N-2}}&\sigma_{_{N-1}}\\
\hdotsfor{7}\\
\sigma_{_{-N+1}} &\cdots&\sigma_{_{-N+p}}&\sigma_{_{-N+p+2}}&\cdots &\sigma_{_{0}}
&\sigma_{_{1}}
\end{pmatrix}}.
\end{equation}
Manipulations similar to those performed in Eq.s (\ref{A20}-\ref{A21}) convert the 
determinant present in (\ref{A23}) into 
\begin{equation}\nonumber
\sum_{1\le j_1,\ldots,j_N\le N}\frac{1}{\epsilon_{j_{1}}^0\epsilon_{j_{2}}^1\ldots
{\epsilon_{j_{N}}}^{N-1}}
{\mathrm \det}
\begin{pmatrix}\nonumber
1  &\cdots&{\epsilon_{j_{1}}}^{p-1}&{\epsilon_{j_{1}}}^{p+1}&\cdots
&{\epsilon_{j_{1}}}^{N-1}&{\epsilon_{j_{1}}}^{N}\\
1  &\cdots&{\epsilon_{j_{2}}}^{p-1}&{\epsilon_{j_{2}}}^{p+1}&\cdots
&{\epsilon_{j_{2}}}^{N-1}&{\epsilon_{j_{2}}}^{N}\\
\hdotsfor{7}\\
1  &\cdots&{\epsilon_{j_{N}}}^{p-1}&{\epsilon_{j_{N}}}^{p+1}&\cdots
&{\epsilon_{j_{N}}}^{N-1}&{\epsilon_{j_{N}}}^{N}\\
\end{pmatrix}.
\end{equation}
Using the property that ${\epsilon_{j}}^N=-\sum_{p=0}^{N-1}a_p{\epsilon_{j}}^p$, 
the above expression becomes
\begin{equation}\nonumber
\sum_{1\le j_1,\ldots,j_N\le N}\frac{-a_p}{\epsilon_{j_{1}}^0\epsilon_{j_{2}}^1\ldots
{\epsilon_{j_{N}}}^{N-1}}
{\mathrm \det}
\begin{pmatrix}\nonumber
1  &\cdots&{\epsilon_{j_{1}}}^{p-1}&{\epsilon_{j_{1}}}^{p+1}&\cdots
&{\epsilon_{j_{1}}}^{N-1}&{\epsilon_{j_{1}}}^{p}\\
1  &\cdots&{\epsilon_{j_{2}}}^{p-1}&{\epsilon_{j_{2}}}^{p+1}&\cdots
&{\epsilon_{j_{2}}}^{N-1}&{\epsilon_{j_{2}}}^{p}\\
\hdotsfor{7}\\
1  &\cdots&{\epsilon_{j_{N}}}^{p-1}&{\epsilon_{j_{N}}}^{p+1}&\cdots
&{\epsilon_{j_{N}}}^{N-1}&{\epsilon_{j_{N}}}^{p}\\
\end{pmatrix},
\end{equation}
and from (\ref{A21}) and (\ref{A23}) one concludes that $q_p=a_p$, $p=0,\ldots,N$. 
In this way, the resolvent of $(\cS)$ coincides with $P_N(z)$. Consequently, the 
$\epsilon_j$s also are unimodular because the assumed non-negativeness of 
$(\cS)$  and \carat's theorem ensure  that the unimodularity is true for the roots 
of resolvent $Q_N(z)$. Thus, the lemma's sufficiency is proven.\#\ac
From the lemma follows, for instance, that the quadratic and cubic 
equations have distinct unimodular roots iff their coefficients are as follows 
\begin{align}
N=2: & \quad a_1=\rho e^{i\phi/2},\quad a_0=e^{i\phi}\quad{\mathrm with}\quad 
0\le \rho <2\ {\mathrm and}\quad \phi\in[0,2\pi)\nonumber\\
N=3:& \quad a_2=\rho e^{i(\phi-\psi)},\ \ a_1=\rho e^{i\psi},\ \ a_0=e^{i\phi}
\quad{\mathrm with}\nonumber \\
\quad&{\mathrm either}\ \ 0\le \rho\le 1,\quad \phi,\psi\in[0,2\pi)\nonumber \\
\quad& {\mathrm or}\ \ 1\le \rho<3,\ \ \phi\in[0,2\pi),\quad \bigl(2\phi -\Phi(\rho)\bigr)
<3\psi<\bigl(2\phi +\Phi(\rho)\bigr)\nonumber
\end{align}
where $\Phi(\rho)\equiv\arccos[(\rho^4+18\rho^2-27)/8\rho^3]$.\ac
An example of Hermitian Toeplitz matrix whose resolvent does not obey the conditions required 
by the lemma because the rank of $(S)$ is smaller than $N$ is given at the end of appendix D.
\section{Properties of the resolvent coefficients}
Given a polynomial equation of degree m 
\begin{equation}\nonumber
P_m(z)\equiv\prod_{j=1}^m (z-\epsilon_j)=\sum_{l=0}^m a_lz^l=0
\end{equation}
with $a_0\ne 0$, we have the interesting property:\ac
{\em the coefficients $a_l$ obey to}
\begin{equation}\label{C0}
{\overline {a_l}}= a_{m-l}/a_0,\ \ l=0,\ldots,m
\end{equation}
{\em iff the roots of the equation are such that 
${\overline {\epsilon_{j}}}=1/{{\epsilon_{i_j}}}$ for $j=1,\ldots,m$ and 
$i_1,\ldots,i_m$ equal to a permutation of $\{1,\ldots,m\}$.}\ac
\# {This condition amounts to say that the polynomial is {\em self-reciprocal}\refr{10}.}  
To prove the necessity one starts from expression (\ref{19}) of $a_l$.  
Taking its complex conjugate and using the assumed property of the roots one finds 
\begin{align}
\overline {a_l}& =(-)^{m-l}\sum_{1\le j_1<\cdots<j_{m-l}\le N} \overline{\epsilon_{j_1}}
\cdots\overline{\epsilon_{j_{m-l}}}\nonumber\\
\quad & =(-)^{m-l}\sum_{1\le j_1<\cdots<j_{m-l}\le N} 
\frac{1}
{\epsilon_{i_{j_1}} \cdots{\epsilon_{i_{j_{m-l}}}}}\nonumber\\
\ &=\frac{(-)^{m-l}}{\epsilon_1\cdots\epsilon_m} 
\sum_{1\le i_1<\cdots<i_{m-l}\le N} \frac{\epsilon_1\cdots\epsilon_m}
{\epsilon_{i_1} \cdots{\epsilon_{i_{m-l}}}}\nonumber\\
\ &=\frac{(-)^{m-l}}{\prod_{j=1}^m \epsilon_j} \sum_{1\le i_1<\cdots<i_{l}\le m} 
\epsilon_{i_1} \cdots{\epsilon_{i_{l}}}=\frac{a_{m-l}}{a_0}.\nonumber
\end{align}
To prove the sufficiency one observes that 
\begin{align}
{\overline {P_m({\overline {z}})}}& =
\prod_{j=1}^m (z-{\overline{\epsilon_{j}}})=
\sum_{j=1}^m {\overline{a_j}}z^j=\sum_{j=1}^m \frac{a_{m-j}}{a_0} z^j\nonumber\\
\quad & =\frac{z^m}{a_0}\sum_{t=0}^m \frac{a_{t}}{z^t}=\frac{z^m}{a_0}
\prod_{j=1}^m (\frac{1}{z}-\epsilon_j).\label{C01}
\end{align}
The previous manipulations require that no root is equal to zero and this is 
ensured by the condition $a_0\ne 0$. With $z=\overline{\epsilon_j}$, whatever $j$ in 
$\{1,\ldots,m\}$, the first product in (\ref{C01}) vanishes. For the second to vanish 
one must have that $1/\overline{\epsilon_j}=\epsilon_{i_j}$ and the property of the roots 
is recovered. In passing it is noted that the property is true also when some roots 
have multiplicity greater than one.\#\ac 
We show now that conditions (\ref{C0}) are obeyed by the $a_l$s defined by 
Eq.s (\ref{26}) and (\ref{27}). 
In fact, setting $l=0$ in Eq.(\ref{26}) one finds 
\begin{equation}\label{C2}
a_0={(-)^{m}\det\bigl(C\overset{\overset{2,\cdots,m+1}{}}
{\underset{1,\cdots,m}{}}\bigr)}/{\mathrm D}=(-)^m e^{i\theta_{_{1}}}
\end{equation}
where the last equality follows putting $n=m+1$, $p=1$ and $q=2$ in (\ref{B7}).  
For the remaining $l$ values we substitute (\ref{26}) in (\ref{C0}) obtaining 
\begin{align}
\ & \overline{{\det\bigl(C\overset{\overset{1,\cdots,l,l+2,\cdots,m+1}{}}
{\underset{1,\cdots,m}{}}\bigr)}}
{\det\bigl(C\overset{\overset{2,\cdots,m+1}{}}
{\underset{1,\cdots,m}{}}\bigr)}\label{C3}\\
\ &=
{\det\bigl(C\overset{\overset{1,\cdots,m-l,m-l+2,\cdots,m+1}{}}
{\underset{1,\cdots,m}{}}\bigr)}
{\det\bigl(C\overset{\overset{1,\cdots,m}{}}
{\underset{1,\cdots,m}{}}\bigr)},\ \quad l=0,\cdots,m\nonumber
\end{align}
where we let $l$ take value $m$ because $a_0\ne 0$. 
Taking $n=(m+1)$, $(j_1,\cdots,j_m)=(1,\cdots,m-l,m-l+2,\cdots,m+1)$ and 
$(i_1,\cdots,i_m)=(1,\cdots,m)$ in (\ref{B3}) 
one finds that 
\begin{equation}\label{c4}
{\det\bigl(C\overset{\overset{1,\cdots,m-l,m-l+2,\cdots,m+1}{}}
{\underset{1,\cdots,m}{}}\bigr)}=
\overline{{\det\bigl(C\overset{\overset{1,\cdots,l,l+2,\cdots,m+1}{}}
{\underset{2,\cdots,m+1}{}}\bigr)}}.
\end{equation}
Using the property that all the {$m\times m$} strictly principal minors 
of $(C)$ coincide, the rhs of (\ref{C3}) becomes %
\begin{equation}\label{C5}
\overline{{\det\bigl(C\overset{\overset{1,\cdots,l,l+2,\cdots,m+1}{}}
{\underset{2,\cdots,m+1}{}}\bigr)}}
{\det\bigl(C\overset{\overset{2,\cdots,m+1}{}}
{\underset{2,\cdots,m+1}{}}\bigr)}.
\end{equation}
From (\ref{B8}) follows that 
\begin{equation}\nonumber
{\det\bigl(C\overset{\overset{2,\cdots,m+1}{}}
{\underset{2,\cdots,m+1}{}}\bigr)}=   
{\det\bigl(\lambda\overset{\overset{1,\cdots,m}{}}
{\underset{2,\cdots,m+1}{}}\bigr)}
{\det\bigl(C\overset{\overset{2,\cdots,m+1}{}}
{\underset{1,\cdots,m}{}}\bigr)}, 
\end{equation}
and 
\begin{equation}\nonumber
{{\det\bigl(C\overset{\overset{1,\cdots,l,l+2,\cdots,m+1}{}}
{\underset{2,\cdots,m+1}{}}\bigr)}}=
{\det\bigl(\lambda\overset{\overset{1,\cdots,m}{}}
{\underset{2,\cdots,m+1}{}}\bigr)}
{{\det\bigl(C\overset{\overset{1,\cdots,l,l+2,\cdots,m+1}{}}
{\underset{1,\cdots,m}{}}\bigr)}}. 
\end{equation}
The substituion of the above two relations in (\ref{C5}) yields 
\begin{equation}\nonumber
\Bigl|{\det\bigl(\lambda\overset{\overset{1,\cdots,m}{}}
{\underset{2,\cdots,m+1}{}}\bigr)}\Bigl|^2 
\overline{{{\det\bigl(C\overset{\overset{1,\cdots,l,l+2,\cdots,m+1}{}}
{\underset{1,\cdots,m}{}}\bigr)}}}{\det\bigl(C\overset{\overset{2,\cdots,m+1}{}}
{\underset{1,\cdots,m}{}}\bigr)}, 
\end{equation}
that coincides with the left hand side of (\ref{C3}) by (\ref{B9}). 
\section{Some properties of Hermitian Toeplitz matrices}
We list here a series of properties obeyed by a square Hermitian Toeplitz matrix 
$(C)$ of order $n$ and partly reported in Ref. [8]. \ac 
{\noindent \bf (a) -} Its elements obey to 
\begin{equation}\label{B1}
{\overline C_{r,s}}=C_{s,r}=c_{r-s}={\overline c_{s-r}},\quad r,s=1,\ldots,n,
\end{equation}
so that all the elements of $(C)$ contained in a line parallel to the main diagonal 
are equal.\ac
{\noindent \bf (b) -}  One has the reflection symmetry with respect to the second 
diagonal formalized by the condition 
\begin{equation}\label{B2}
C_{r,s}=c_{n+1-s,n+1-r}.
\end{equation}
{\noindent \bf (c) -} All the $(m\times m)$ principal minors of $(C)$, 
whatever the considered rows (and columns), are identical. \ac
\# The property is a consequence of {\bf (a)}.\#\ac
{\noindent \bf (d) -} For any choice of $m$ rows  $(1\le i_1<\cdots<i_m\le n)$ 
and $m$ columns $(1\le j_1<\cdots<j_m\le n)$ it results
\begin{equation}\label{B3}
\det\bigl(C
\overset{\overset{j_1,\cdots,j_m}{}}{\underset{i_1,\cdots,i_m}{}}\bigr)=
\det\bigl(C^T
\overset{\overset{(n+1-j_m),\cdots,(n+1-j_1)}{}}
{\underset{(n+1-i_m),\cdots,(n+1-i_1)}{}}\bigr)=
\overline{
\det\bigl(C
\overset{\overset{(n+1-j_m),\cdots,(n+1-j_1)}{}}
{\underset{(n+1-i_m),\cdots,(n+1-i_1)}{}}\bigr)}.
\end{equation}
\# The first equality, where $(C^T)$ denotes the transposed of $(C)$, 
follows from property {\bf (b}) and the second from the 
Hermiticity of $(C)$. \# \ac
The following properties, that we think to be original, hold only true for 
Hermitian Toeplitz matrices having their rank equal to the principal one.\ac
{\noindent \bf (e) -} If the principal rank of a Hermitian Toeplitz matrix $(C)$ 
is equal to the rank $m (\le n)$ of $(C)$, any $(m\times m)$ strictly principal minor of 
$(C)$ is non-singular. \ac
\# The property immediately follows from the definition of "principal rank"  of 
a matrix, reported above Eq. (\ref{8}), and ({\bf c})\#.\ac
{\noindent \bf (f) -} For any $(n\times n)$ Hermitian Toeplitz matrix of rank equal 
to its principal rank $m(\le n)$, the determinant of  any of its minors formed 
by $m$ {\em subsequent} rows and $m$ {\em subsequent} columns is simply related 
by a phase factor to the determinant of the (strictly) principal minor contained in the 
considered rows or columns, \ie\ 
\begin{align} 
\det\bigl(C
\overset{\overset{(q+1),\cdots,q+m}{}}{\underset{p+1,\cdots,p+m}{}}\bigr)
&=e^{i\theta_{p-q}}
\det\bigl(C
\overset{\overset{p+1,\cdots,p+m}{}}
{\underset{p+1,\cdots,p+m}{}}\bigr),\label{B6}\\ 
\det\bigl(C
\overset{\overset{(q+1),\cdots,q+m}{}}{\underset{p+1,\cdots,p+m}{}}\bigr)
& =e^{i\theta_{p-q}}
\det\bigl(C
\overset{\overset{q+1,\cdots,q+m}{}}
{\underset{q+1,\cdots,q+m}{}}\bigr),\nonumber\\
\ & \theta_{_{p-q}}\in {\mathbb R}, \quad p,q=0,1,\ldots,n-m.\nonumber
\end{align}
\# Clearly, if the first of the above two equalities is true the second also is true 
because of ({\bf c}). To prove the first of equalities (\ref{B6}) one observes that 
({\bf d}) implies that any $m$ distinct rows of $(C)$ can be written as linear 
combinations of $m$ other distinct rows (see, {\em e.g.,} Ref. [18], Chap. III). 
Hence rows $(p+1),\ldots,(p+m)$ can be expressend in terms of rows $(q+1),\ldots,
(q+m)$ as 
\begin{equation}\label{B7}
C_{r,s}=\sum_{t=q+1}^{q+m}\lambda_{r,t}C_{t,s}, \quad r=(p+1),\ldots,(p+m),\quad 
s=1,\ldots,n,
\end{equation}
where the $\lambda_{r,t}$ are suitable numerical coefficients. From these relations 
follows that 
\begin{equation}\label{B7bis}
\det\bigl(C
\overset{\overset{p+1,\cdots,p+m}{}}
{\underset{p+1,\cdots,p+m}{}}\bigr) = 
\det\bigl(\lambda
\overset{\overset{q+1,\cdots,q+m}{}}
{\underset{p+1,\cdots,p+m}{}}\bigr)
\det\bigl(C
\overset{\overset{p+1,\cdots,p+m}{}}
{\underset{q+1,\cdots,q+m}{}}\bigr).
\end{equation}
Due to ({\bf e})  the left hand side of (\ref{B7bis}) is different from zero so that 
both factors on the rhs are different from zero. 
The complex conjugation of (\ref{B7bis}), by the Hermiticity of $(C)$, yields 
\begin{equation}\label{B8}
\det\bigl(C
\overset{\overset{p+1,\cdots,p+m}{}}
{\underset{p+1,\cdots,p+m}{}}\bigr) = 
\overline{\det\bigl(\lambda
\overset{\overset{q+1,\cdots,q+m}{}}
{\underset{p+1,\cdots,p+m}{}}\bigr)}
\det\bigl(C
\overset{\overset{q+1,\cdots,q+m}{}}
{\underset{p+1,\cdots,p+m}{}}\bigr).
\end{equation}
From Eq. (\ref{B7}) also follows that 
\begin{align}
\det\bigl(C
\overset{\overset{q+1,\cdots,q+m}{}}
{\underset{p+1,\cdots,p+m}{}}\bigr) &= 
\det\bigl(\lambda
\overset{\overset{q+1,\cdots,q+m}{}}
{\underset{p+1,\cdots,p+m}{}}\bigr)
\det\bigl(C
\overset{\overset{q+1,\cdots,q+m}{}}
{\underset{q+1,\cdots,q+m}{}}\bigr)\nonumber\\
\ &=\det\bigl(\lambda
\overset{\overset{q+1,\cdots,q+m}{}}
{\underset{p+1,\cdots,p+m}{}}\bigr)
\det\bigl(C
\overset{\overset{p+1,\cdots,p+m}{}} 
{\underset{p+1,\cdots,p+m}{}}\bigr),\nonumber
\end{align}
where the last equality follows from ({\bf c}). The substituion of the 
last equality in Eq. (\ref{B7bis}) and the fact that,  by assumption, 
$\det\bigl(C\overset{\overset{p+1,\cdots,p+m}{}}
{\underset{p+1,\cdots,p+m}{}}\bigr)\ne 0$ imply that 
\begin{equation}\label{B9}
\Bigl|\det\bigl(\lambda
\overset{\overset{q+1,\cdots,q+m}{}}
{\underset{p+1,\cdots,p+m}{}}\bigr)\Bigl|^2=1\quad p,q=0,\ldots, n-m, 
\end{equation} 
and Eq. (\ref{B6}) is proven. That the phase factor depends on $p-q$ instead of 
$(p,q)$ follows from the fact that the two determinants present in (\ref{B6}) 
do not change with the two substitutions $p\to p+1$ and $q\to q+1$ owing 
to (\ref{A2}).\#\ac
An immediate consequence of ({\bf f}) is the property that\ac
{\noindent \bf (g) -} any $(m\times m)$ minor formed by $m$ subsequent rows and $m$ 
subsequent columns of a Hermitian Toeplitz matrix with rank equal to its principal 
rank $m$ is non-singular. 
\section{Numerical examples}
To illustrate the application of the results discussed above we report three numerical 
examples.\ac
{\bf 1 -} The first shows a case where it is impossible to express a set of $c_j$s in 
terms of positive and negative $\rho_j$s.  Assume that 
$c_1=0,\ c_2=0,\ c_3=1$. The corresponding matrix $(\hat C)$ has eigenvalues equal 
to $-1,\ 0,\ 0,\ 1$. The matrix $(C_1)$, obtained by setting $c_0=1$, is 
semi-positive definite with rank 3 and eigenvalues equal to $0,\ 1,\ 1,\ 2$. 
The solutions are: $\rho_1=\rho_2=\rho_3=1/3$ and $\epsilon_1=1$, 
$\epsilon_2=-e^{i\pi/3}$, $\epsilon_3=e^{i2\pi/3}$. 
The matrix $(C_2)$,  obtained by setting $c_0=-1$, is semi-negative definite with 
rank equal to 3 and eigenvalues equal to $-2,\ -1,\ -1,\ 0$. 
The solutions are: $\rho_1=\rho_2=\rho_3=-1/3$ and $\epsilon_1=-1,\ 
\epsilon_2=e^{i\pi/3},\ \epsilon_3=-e^{i2\pi/3}$. 
Finally, the matrix $(C_3)$ obtained by setting $c_0=0$ coincides with $(\hat C)$. 
It is non-definite, has rank equal to 2 and principal rank equal to 0. For this 
reason it is impossible to write 0, 0, 0 and 1 in the form (1) with $m=2$ as it 
is easily checked. \ac
{\bf 2 -} The second example considers the case where $c_1=1,\ c_2=0,\ c_3=1$. The 
eigenvalues of the associated matrix $(\hat C)$ are $-2,\ 0,\ 0, 2$. Setting 
$c_0=0$ the resulting $(C)$ matrix coincides with $(\hat C)$. It is non-definite, 
its rank is two and equal to its principal rank value. The generalized \carat's 
theorem applies and the solution is $\epsilon_1=1$, $\epsilon_2=-1$, 
$\rho_1=1/2$ and $\rho_2=-1/2$. With these values one easily checks that  
$c_0=\rho_1+\rho_2=0$,  $c_1=\rho_1\epsilon_1+\rho_2\epsilon_2=1$, 
$c_2=\rho_1\epsilon_1^2+\rho_2\epsilon_2^2=0$ and $c_3=\rho_1\epsilon_1^3+
\rho_2\epsilon_2^3=1$.\ac 
Setting $c_0=2$ the resulting $(C)$ is non-negative defined with rank 3. The 
solution is: $\epsilon_1=1$, $\epsilon_2=i$, $\epsilon_3=-i$, 
$\rho_1=1$, $\rho_2=1/2$ and $\rho_3=1/2$. The last choice $c_0=-2$ defines a 
non-positive defined $(C)$ matrix with rank equal to 3 with 
solution: $\epsilon_1=-1$, $\epsilon_2=i$, $\epsilon_3=-i$, 
 $\rho_1=-1$, $\rho_2=-1/2$ and $\rho_3=-1/2$.\ac
{\bf 3 -} The last example corresponds to have $c_0=\delta_0$, 
$c_1=\delta_0+i\delta_1 e^{i\varphi}$ and $c_2=(\delta_0+2i\delta_1)e^{2i\varphi}$ 
with $\delta_0$, $\delta_1$ and $\varphi$ reals. The eigenvalues are 
0 and $(3\delta_0/2)(1\pm \sqrt{1+8\delta_1^2/(3\delta_0^2)})$. The Hermitian  matrix $(C)$ 
is indefinite and has rank 2 under the assumption that no further eigenvalue 
is equal to zero.  The resolvent equation is $P(z)= z^2-2e^{i\varphi}z+e^{2i\varphi}=0$ so that 
its coefficients $a_0=e^{2i\varphi}$, $a_1=-2e^{i\varphi}$ and $a_2=1$ obey Eq. (\ref{A1bis}). 
From (39) one immediately finds that $\sigma_0=2$, $\sigma_1=2e^{i\varphi}$ and 
$\sigma_2=2e^{2i\varphi}$ and the resulting matrix $(\cS)$ has rank 1. 
Hence the previous resolvent equation has a unimodular root with multiplicity 2: 
in fact, $P(z)=(z-e^{i\varphi})^2=0$. 
According to our analysis, $(C)$ cannot be written as $(\cV)(\Delta)(\cV)^{\dag}$ 
with $(\Delta)$ equal to a diagonal matrix. In this example, however, the roots are unimodular 
and $(C)$ factorizes in the Ellis-Lay form\refr{9} as 
\begin{align}\nonumber
\ &\begin{pmatrix}
\delta{_0},  &(\delta_0+i\delta{_1})e^{i\varphi},&(\delta_0+2i\delta{_1})e^{2i\varphi}\\
(\delta_0-i\delta{_1})e^{-i\varphi},&\delta{_0},&(\delta_0+i\delta{_1})e^{i\varphi}\\
(\delta_0-2i\delta{_1})e^{-2i\varphi},&(\delta_0-i\delta{_1})e^{-i\varphi},&\delta_0
\end{pmatrix}=  \\
\ & \quad \nonumber\\
\ & \begin{pmatrix}
 1,&0\\
e^{-i\varphi},&e^{-i\varphi}\\
e^{-2i\varphi},&2e^{-2i\varphi}\\
\end{pmatrix} 
\begin{pmatrix}
\delta_0, &i\delta_1\\
-i\delta_1,&  0
\end{pmatrix}
\begin{pmatrix}
1,& e^{i\varphi},&e^{2i\varphi}\\
0,& e^{i\varphi},&2e^{2i\varphi}
\end{pmatrix}.
\end{align}
%As already noted in the introductory section, this more general factorization finds 
%no application to x-ray/neutron scattering or signal theory. 

\newpage

\leftline{\bf References}\noindent
\begin{itemize}
\item[\ ${}^{1}$] {C. Carath\'eodory, {\em  Rend. Circ. Matem. Palermo} {\bf 32}, 
           193 (1911).}
\item[\ ${}^{2}$] {A. Cervellino and S. Ciccariello, {\em Riv. Nuovo Cimento} {\bf 19/8}, 
           1 (1996).}
\item[\ ${}^{3}$] {V.F. Pisarenko, {\em Geophys. J. R. Astr. Soc.} {\bf 33}, 346 (1973).}
\item[\ ${}^{4}$] {G. Beylkin and L. Monzon, {\em Appl. Comp. Harmon. Anal.} 
{\bf 12}, 332 (2002).}
\item[\ ${}^{5}$] {L.A. Feigin and D.I. Svergun, {\em Structure Analysis by Small-Angle X-ray 
and Neutron Scattering}, (Plenum Press, New York, 1987).}
\item[\ ${}^{6}$] {A. Cervellino and S. Ciccariello, {\em Acta Crystall. A}{\bf 61}, 
           494 (2005).}
\item[\ ${}^{7}$] {U. Grenander and G. Szeg\"o, {\em Toeplitz forms and their 
applications}, (Chelsea, New York, 1984) p. 56-60.}
\item[\ ${}^{8}$] {I.S. Iohvidov, {\em Hankel and Toeplitz matrices and forms: algebraic theory}, 
(Birkhauser, Boston, 1982).}
\item[\ ${}^{9}$] {R.L. Ellis and D.C. Lay, 
%{\em Factorization of finite rank Hankel and Toeplitz matrices}, 
{\em Linear Algebra Appl.} {\bf 173}, 19 (1992).}
\item[\ ${}^{10}$]  {W. Chen, {\em J. Math. Anal. Appl.} {\bf 190}, 714 (1995).}
\item[\ ${}^{11}$] {J.M. McNamee, {\em J. Comp. Appl. Math.} {\bf 110}, 305 (1999).}
\item[\ ${}^{12}$] {F.R. Gantmacher, {\em Th\'eorie des matrices}, (Dunod, Paris, 1966), 
Vol. I.}
\item[\ ${}^{13}$] {M.L. Mehta, {\em Matrix Theory: Selected topics and useful 
results}, (Les Ulis, Les Editions de Physique, 1989)}
\item[\ ${}^{14}$] This is also evident from (\ref{15}): by construction 
the rows of $(V)$ span the range space of $(C)$ and 
$(C)$ has rank at most equal to $m$.
\item[\ ${}^{15}$] { For the case of Hermitian non-negative Toeplitz matrices considered 
by Grenander and G. Szeg\"o,  their proof shows that the principal rank of these matrices 
is equal to their rank. An alternative proof was given by Goedkoop\refr{16} introducing a 
finite dimensional Hilbert space (see also Ref. [2]). Thus, for this kind of matrices, 
one has the interesting property that the rank is obtained by considering the 
strictly principal minors of increasing order till finding a singular minor. If 
the latter's order is $m+1$ the rank of the matrix is $m$.} 
\item[\ ${}^{16}$] J.A. Goedkoop, {\em Acta Crystall.} {\bf 3}, 374 (1950).
\item[\ ${}^{17}$] E.W. Weisstein, {\em Sylvester's Inertia Law}, MathWorld--A 
Wolfram Web: 
{\texttt{http://mathworld.wolfram.com/CongruenceTransformation.html}}.
\item[\ ${}^{18}$] {R. Caccioppoli, {\em Lezioni di Analisi 
Matematica}, (Liguori, Naples, 1956), Vol. I.}
\item[\ ${}^{19}$] {D. Cox, J. Little and D. O'Shea, {\em Using algebraic geometry}, 
(Springer, Berlin, 1998).}
%%%%\item[\ ${}^{17}$] This example was brought to our attention by the referee to 
%%%%alert us that a flaw was present in our earlier proof of properties $(d)-(g)$.

\end{itemize}
\end{document}